\documentclass[twocolumn,showpacs,superscriptaddress,pra]{revtex4}

\usepackage{amsmath, amsthm, amssymb}
\usepackage{graphicx}
\usepackage{dcolumn}
\usepackage{bm}
\usepackage{color}

\begin{document}
\title{Robustness of Entanglement as a Resource}
\author{Rafael Chaves}
\affiliation{Instituto de F\'\i sica, Universidade Federal do Rio de
Janeiro. Caixa Postal 68528, 21941-972 Rio de Janeiro, RJ, Brasil}
\author{Luiz Davidovich}
\affiliation{Instituto de F\'\i sica, Universidade Federal do Rio de
Janeiro. Caixa Postal 68528, 21941-972 Rio de Janeiro, RJ, Brasil}
\date{\today}

\begin{abstract}
The robustness of multipartite entanglement of systems undergoing
decoherence is of central importance to the area of quantum
information. Its characterization depends however on the measure
used to quantify entanglement and on how one partitions the
system. Here we show that the unambiguous assessment of the
robustness of multipartite entanglement is obtained by considering
the loss of functionality in terms of two communication tasks,
namely the splitting of information between many parties and the
teleportation of states.

\end{abstract}

\pacs{03.67.-a, 03.67.Mn, 03.65.Yz} \maketitle

\section{Introduction.} \label{intro}

The fast development of quantum information in the last two decades
has strengthened the notion that entanglement is not only a
fundamental concept in quantum mechanics but also a resource for
processing and transmitting information. For bipartite systems, the
relationship between entanglement measures and the use of
entanglement for specific tasks like teleportation
\cite{teleportation} and quantum key distribution \cite{QKD} is well
understood. However, the exact role of multipartite entanglement in
communication protocols and in quantum computation is still an open
question. In the multipartite context, many different measures and
kinds of entanglement are possible \cite{Horodecki entanglement
review}. But it is still unclear what is the relevance of these
measures for quantifying the ability of performing different tasks,
and what is the role of each kind of entanglement in surpassing the
classical gain for a given protocol.

We show in this paper that these questions are particularly relevant
for open systems. Indeed, different kinds of entanglement can behave
quite differently under decoherence. Once the resource for a given
protocol is (non)robust against perturbations, this implies that the
protocol itself should also be (non)robust. In fact, we show here
that the comparison between the effect of the environment on the
many kinds of entanglement for a multipartite system, used as a
resource for a given task,  and the effect of decoherence on how
well the task is achieved indicates which type of entanglement is
relevant for accomplishing it.

The dynamics of entanglement, for a wide class of entangled
states, have been extensively analyzed
\cite{suddendeath,simonkempe,dur,aolita,cavalcanti,andre,montakhab,konrad}
and experimentally demonstrated \cite{esdexp,osvaldo}. The
non-unitary evolution depends intrinsically on the
system-environment dynamics and the quantum entangled state under
consideration. Besides a more general treatment restricted to a
two-qubit case given in \cite{konrad,osvaldo}, all analysis have
restricted themselves to particular initial entangled states
evolving under some specific dynamics
\cite{suddendeath,simonkempe,dur,aolita,andre,montakhab,cavalcanti}.
These studies have shown that the dynamics of entanglement is
quite distinct from the dynamics of populations and coherences. In
particular, entanglement may vanish at a finite time, much before
the asymptotic decay of the coherences
\cite{suddendeath,simonkempe,dur,aolita}.

For multipartite systems the problem gets more involved since
there are inequivalent classes of entangled states \cite{cirac},
that is, states that are not connected by local operations and
classical communication (LOCC's). States that can be obtained from
each other through LOCC's belong to the same class of states,
meaning that these states are equivalent resources for a large
class of tasks in quantum information, those that are invariant
under LOCC's, like teleportation \cite{teleportation} and
distillation \cite{distillation}.

Many of the investigations on the open-system dynamics of
entanglement refer to $GHZ$ states
\cite{simonkempe,dur,aolita,andre,montakhab} 
\begin{equation}
\left\vert GHZ_{N}\right\rangle =\alpha\left\vert 0\right\rangle ^{\otimes N}+\beta\left\vert 1\right\rangle ^{\otimes N}, \label{ghz}
\end{equation}
subject to the normalization condition $\left\vert \alpha\right\vert
^{2}+\left\vert \beta\right\vert ^{2}=1$.

Another class of entangled states, the $W$ states 
\begin{equation}
\left\vert W_{N}\right\rangle =\frac{1}{\sqrt{N}}\left(  \left\vert 00..01\right\rangle +\left\vert 01..00\right\rangle +..+\left\vert 00..01\right\rangle \right)  , \label{wstate}
\end{equation}
were introduced in \cite{cirac}, where it was shown that they are
inequivalent to the $GHZ$ states.

It is easy to see that the entanglement of $W$ states is more robust
than that of $GHZ$ states upon loss of  a particle \cite{cirac}.
Given this symptomatic entanglement robustness of $W$ states, it is
natural to ask which class of states, $GHZ$ or $W$, is more robust
in a more general situation where each part of a multipartite state
is undergoing decoherence. The entanglement dynamics analysis of $W$
and $GHZ$ states is relevant on its own, since these states are
resources for quantum information protocols as quantum
teleportation, dense coding and secure distribution of quantum keys
\cite{GHZuse,wuse}, and have been experimentally produced in many
kinds of physical systems including atoms in cavities, trapped ions
and entangled photons \cite{ghzexp,wexp}.

The simple form of $N$-partite $GHZ$ states has allowed several
analytical derivations of the associated entanglement dynamics
\cite{simonkempe,dur,aolita}. An interesting conclusion was drawn
for multipartite $GHZ$ states subjected to local environments
\cite{aolita}: the time at which bipartite (and multipartite)
entanglement becomes arbitrarily small can occur well before the
disentanglement time; this time difference strongly increases with
the number of particles. This implies that, for $GHZ$ states, the
time for which entanglement becomes arbitrarily small better
characterizes the entanglement robustness than the disentanglement
time. On the other hand, the global entanglement of $W$ states was
shown to be more robust: the decay rate is size independent for
dephasing and zero temperature environments
\cite{andre,montakhab}. This previous work motivates the question
on how other types of entanglement, defined between different
kinds of partitions of the state, behave for $W$ states under
decoherence. The physical meaning of these different types of
entanglement and the corresponding measures is also relevant, the
important question being how do they relate to possible
applications in computation and communication.

In this paper we analyze the original $W$ state (\ref{wstate}) and
related states, referred to as $W$-like states,
\begin{equation}
\left\vert \widetilde{W}_{N}\right\rangle =a_{1}\left\vert 10..00\right\rangle +a_{2}\left\vert 01..00\right\rangle +..+a_{N}\left\vert 00..01\right\rangle ,
\label{wlike}
\end{equation}
\begin{equation}
\left\vert {W}_{N}^{0}\right\rangle =\alpha\left\vert {W}_{N}\right\rangle +\beta\left\vert 0\right\rangle ^{\otimes N},
\label{wlike0a}
\end{equation}
\begin{equation}
\left\vert \widetilde{W}_{N}^{0}\right\rangle =\alpha\left\vert \widetilde {W}_{N}\right\rangle +\beta\left\vert 0\right\rangle ^{\otimes N}\,.
\label{wlike0b}
\end{equation}
We show that, when considering bipartite entanglement, these states
display the same kind of non-robustness than the $GHZ$ states. For
two kinds of decoherence, we derive an analytical expression, valid
for any entanglement measure, which can be expressed as the
entanglement of that of considerably smaller subsystems, in the same
spirit as in \cite{cavalcanti}. In the dephasing case, the bipartite
and global entanglement in $W$ states remain both very robust when
the number of particles increases, the disentanglement  time being
equivalent in this case to the time at which entanglement becomes
very small. Under the amplitude damping channel, these two time
scales become quite distinct: the bipartite entanglement as
quantified by the negativity \cite{vidal}, for a sufficiently large
$N$, decays below an arbitrarily small value much before it
vanishes, in spite of the fact that the global entanglement is size
independent. This clearly shows how differently can distinct kinds
of entanglement behave under decoherence. Adopting a more practical
point of view and explicitly considering as tasks the teleportation
\cite{teleportation} and the splitting \cite{splitting} of quantum
information, we show that the dynamics of fidelity for each of these
protocols, when using as a resource an entangled state undergoing
decoherence, can be identified with the decay of a specific kind of
entanglement measure, thus allowing the identification of each of
these measures with a well-defined task.

Our paper is organized as follows. In Sec. \ref{deco}, we
introduce the decoherence maps to be considered in this article.
In Sec. \ref{em}, we discuss the entanglement quantifiers that are
used to characterize the entanglement of multipartite states. In
Sec. \ref{ed}, we introduce a method that allows for the
calculation of the entanglement corresponding to a given
bipartition of a $W$ or $W$-like multipartite state through the
calculation of the entanglement of a two-qubit system, which
greatly simplifies the analysis of the dynamics of entanglement
for this kind of state. This technique is used in Sec. \ref{bg} to
compare the dynamics of global and bipartite quantifiers of
entanglement, which have been proposed in the literature. The
quite different features under decoherence of these quantifiers is
shown in Sec. \ref{teleport} to have a direct relation to the
robustness of inequivalent communication tasks. In Sec. \ref{conc}
we summarize our conclusions. Detailed calculations are given in
the Appendix.

\section{Decoherence models.}\label{deco}

The dynamics of a system interacting with an environment can be
described in the Kraus form \cite{nielsenchuang}, so that the
density operator of the evolved system is given by $\rho=\sum_j
E_j\rho_0E_j^\dagger$, where $E_j$ are positive-operator valued
measures (POVM's) satisfying $\sum_j E_j^\dagger E_j=1$, and
$\rho_0$ is the initial state of the system. We analyze here the
entanglement dynamics of the states (\ref{wstate}), (\ref{wlike}),
(\ref{wlike0a}) and (\ref{wlike0b}), evolving under the action of
two paradigmatic noisy channels: dephasing and amplitude damping.
The states are described in the computational basis $\left\{
\left\vert 0\right\rangle ,\left\vert 1\right\rangle \right\} $. We
assume that the qubits  have identical interactions with their own
individual environments, and neglect the mutual interaction between
qubits. The dynamics of the $i$-th qubit, $1\leq i\leq N$, is then
described by a completely positive trace-preserving map (or channel)
$\mathcal{E}_{i}$, so that
 $\rho=\mathcal{E}_{i}\rho_{0}$, while the evolution of the $N$-qubit system is given by the composition of all $N$ individual maps: $\rho\equiv \mathcal{E}_{1}\mathcal{E}_{2}\
...\ \mathcal{E}_{N}\rho_{0}$. The assumption of mutually
independent and identical environments is well justified whenever
the separation between the particles is much larger than a typical
length associated with the environment (like a resonant wavelength
of an electromagnetic environment), so that collective effects do
not need to be taken into account.

The phase damping (or dephasing) channel ($D$) represents the
situation in which there is loss of coherence with probability
$p$, but without any energy exchange. It is defined as 
\begin{equation}
\varepsilon_{i}^{D}\rho_{i}=\left(  1-p\right)  \rho_{i}+p\left(  \left\vert 0\right\rangle \left\langle 0\right\vert \rho_{i}\left\vert 0\right\rangle \left\langle 0\right\vert +\left\vert 1\right\rangle \left\langle 1\right\vert \rho_{i}\left\vert 1\right\rangle \left\langle 1\right\vert \right)  .
\label{dephasing}
\end{equation}
The application of this map in the computational basis $\left\{
\left\vert 0\right\rangle ,\left\vert 1\right\rangle \right\}  $
clearly does not affect the populations, but all the original
coherences $\left\vert i\right\rangle \left\langle j\right\vert $
get reduced by a decay factor $(1-p)$.

The amplitude-damping channel ($AD$) is given, in the Born-Markov approximation, via its Kraus representation, as:%
\begin{equation}
\mathcal{E}_{i}^{AD}\rho_{i}=E_{0}\rho_{i}E_{0}^{\dagger}+E_{1}\rho_{i} E_{1}^{\dagger}, \label{amplitude damping}
\end{equation}
with $E_{0}\equiv\left\vert 0\right\rangle \left\langle 0\right\vert
+\sqrt{1-p}\left\vert 1\right\rangle \left\langle 1\right\vert $,
$E_{1} \equiv\sqrt{p}\left\vert 0\right\rangle \left\langle
1\right\vert $. Here $|1\rangle$ stands for the excited state and
$|0\rangle$ for the ground state of a two-level system. In this
case, the population of the excited state is reduced by a factor
$1-p$, while the population of the ground state is equal to the
initial population plus the contribution coming from the upper
state, which is equal to $p$ multiplied by the initial population of
that state. The coherences, on the other hand, are reduced by the
factor $\sqrt{1-p}$.

The parameter $p$ in channels (\ref{dephasing}) and
(\ref{amplitude damping}) is a convenient parametrization of time:
$p = 0$ refers to the initial time $t=0$ and $p= 1$ refers to the
asymptotic t$\rightarrow\infty$ limit. In the Markovian
approximation, and for the amplitude channel, one has
$p=1-\exp(-\gamma t)$, where $\gamma$ is the decay rate of the
excited state. This corresponds to the well-known Weisskopf-Wigner
approximation. It should be remarked that these two maps, when
acting on $GHZ$ state (\ref{ghz}) and $W$-like states
(\ref{wstate}), (\ref{wlike}), (\ref{wlike0a}), and
(\ref{wlike0b}), do not create new coherences, but for the
amplitude-damping channel new diagonal terms (populations) may be
created.

\section{Entanglement Measures}\label{em}

To investigate the entanglement features of the states here
considered and the relation of these features with different
communication tasks, we compare different quantifiers of
entanglement. In particular, we evaluate explicitly the dynamics
of the negativity \cite{vidal} and concurrence \cite{concurrence},
measures of bipartite entanglement in a given bipartition of a
multipartite state, and of the Meyer-Wallach measure of global
entanglement ($MW$) \cite{meyerwallach}. We also compare the
negativity, concurrence and the $MW$ measure with the generalized
concurrence \cite{rungta}, a measure of genuine multipartite
entanglement that had its dynamics in the $W$ state (\ref{wstate})
numerically calculated in \cite{andre}.

The negativity, given a bipartition $\{A\}:\{B\}$ of a multipartite
state, is defined as the absolute value of the sum of the negative
eigenvalues of the partially transposed ($PT$) density matrix
$\rho_{AB}^{T_{A}}$ \cite{pereshorodecki}, which can be defined in
terms of the trace norm $\left\Vert \rho_{AB}^{T_{A}}\right\Vert $,
the sum of moduli of the eigenvalues of $\rho_{AB}^{T_{A}}$, as
\cite{vidal}
\begin{equation}
\mathcal{N}\left(  \rho\right)  =\frac{\left\Vert \rho_{AB}^{T_{A}}\right\Vert -1}{2}.\label{negdefinition}
\end{equation}
In general, the negativity fails to quantify entanglement of some
entangled states (those with positive partial transposition) in
dimensions higher than six \cite{pereshorodecki}. However, for the
states considered here, under the influence of the two maps
considered in the previous section, dephasing and amplitude
damping, the negativity vanishes only when the state is a
separable one. So, in these cases, the negativity brings all the
relevant information about the separability in bipartitions of the
states (i.e., null negativity means separability in the
corresponding partition). In particular, as we show in the
following section, the bipartite entanglement problem for $W$-like
states (\ref{wstate}), (\ref{wlike}), (\ref{wlike0a}), and
(\ref{wlike0b}), can always be reduced to two qubits, a situation
where the negativity is an unambiguous entanglement quantifier.

The concurrence \cite{concurrence} for a given bipartition
$\{A\}:\{B\}$ of a multipartite pure state $\left\vert
\psi\right\rangle$ is
\begin{equation}
C\left(  \left\vert \psi\right\rangle \right)  = \sqrt{2\left(
1-{\rm tr}\rho_{A}^{2}\right)} , \label{concqudit}
\end{equation}
being $\rho_{A}= \rm{tr}_{B} \left( \rho_{AB} \right)$. This measure
can be extended over the mixed states $ \rho = \sum_{i}
p_{i}\left\vert \Psi_{i}\right\rangle \left\langle
\Psi_{i}\right\vert $ by virtue of a convex roof construction
\cite{convexroof}
\begin{equation}
C \left(  \rho \right)  = \inf_{\left\{  p_{i},\left\vert
\Psi_{i}\right\rangle \right\}  } {\displaystyle\sum\limits_{i}}
p_{i}C\left(  \left\vert \Psi_{i}\right\rangle \right),
\end{equation}
an optimization that can be analytically evaluated for two-qubit
states, the minimum in this case is obtained by
\begin{equation}
\label{concmixed}
 C\left(  \rho\right)  =\max\left\{  0,\sqrt{\lambda_{1}}-\sqrt{\lambda_{2}%
}-\sqrt{\lambda_{3}}-\sqrt{\lambda_{4}}\right\},
\end{equation}
with $\lambda_{i}$ the eigenvalues, $\lambda_{1}$ denoting the
largest among them, of the matrix $\rho\left(
\sigma_{y}\otimes\sigma_{y}\right) \rho^{\ast}\left( \sigma
_{y}\otimes\sigma_{y}\right)$, where the conjugation is taken with
respect to the computational basis $\left\{ \left\vert
0\right\rangle ,\left\vert 1\right\rangle \right\} $.

For a pure state $\left\vert \psi\right\rangle $ of $N$ qubits and a
partition $\mathcal{A}|\mathcal{B}$, the corresponding nonlocal
information $S_{\mathcal{A}|\mathcal{B}}$ between $\mathcal{A}$ and
$\mathcal{B}$ is distributed among different kinds of contributions
\cite{guo} 
\begin{equation}
S_{\mathcal{A}|\mathcal{B}}= {\displaystyle\sum\limits_{k=2}^{N}}
I_{i_{1}i_{2}\ldots i_{k}}, \label{biasmulti}
\end{equation}
where the sum is taken over all possible combinations of indices,
such that $i_{1},i_{2},\ldots,i_{k}$ are not in the same set
$\mathcal{A}$ or $\mathcal{B}$, and $I_{i_{1}i_{2}\ldots i_{k}}$ is
some appropriate measure of nonlocal information shared among all
the $k$ qubits. The quantity $S_{\mathcal{A}|\mathcal{B}}$ can be taken \cite{guo} as the mutual information%
\begin{equation}\label{sab}
S_{\mathcal{A}|\mathcal{B}}=S_{\mathcal{A}}+S_{\mathcal{B}}-S_{\mathcal{AB}},
\end{equation}
with $S_{\mathcal{Y}}=\left(  1-{\rm
tr}\rho_{\mathcal{Y}}^{2}\right)$ being the linear entropy. Since
$\left\vert \psi\right\rangle $ is pure, $S_{\mathcal{AB}}=0$, the
entropies of the two parts ${\cal A}$ and ${\cal B}$ are
identical, and therefore $S_{\mathcal{A}|\mathcal{B}}=2\left(
1-{\rm tr}\rho_{\mathcal{A}}^{2}\right)  $. For example, for
two-qubit pure states, $I_{12}=\tau_{12}$, where $\tau_{12}$ is
the square of the concurrence \cite{concurrence} and is called the
$2$-tangle. For three-qubit pure states, the corresponding
nonlocal information between $1$ and $23$ can be written as
$S_{1|23}=I_{12}+I_{13}+I_{123}$ with $I_{12}=\tau_{12}$,
$I_{13}=\tau_{13}$, and $I_{123}=\tau_{123}$, where the $3$-tangle
$\tau_{123}$ has been shown to be a well-defined measure of
genuine three-qubit entanglement \cite{3tangle}.

The $MW$ measure for entanglement \cite{meyerwallach} for a pure
state $\left\vert \psi\right\rangle $ of $N$ qubits can be expressed
as \cite{brennen} 
\begin{equation}
E_{MW}\left(  \left\vert \psi\right\rangle \right)  =\frac{1}{N}
{\displaystyle\sum\limits_{i=1}^{N}} 2\left(
1-{\rm tr}\rho_{i}^{2}\right), \label{mwbybrennen}
\end{equation}
an average over the entanglement (square of the concurrence) of each
qubit with the rest of the system. From Eqs.~ (\ref{biasmulti}) and
(\ref{sab}) we can see that the $MW$ measure can be distributed
among different kinds of nonlocal information
\begin{eqnarray}
E_{MW}\left(  \left\vert \psi\right\rangle \right)  &=&\frac{1}{N}\left(  2\sum_{i_{1}<i_{2}}%
I_{i_{1}i_{2}}+\ldots\right.\nonumber\\
&+&\left.N\sum_{i_{1}<i_{2}\ldots<i_{N}}I_{i_{1}i_{2}\ldots
i_{N}}\right)  .
\end{eqnarray}
From this last equation it is clear that the $MW$ measure depends
on the different forms of quantum correlations present in the
entangled state and in fact it was originally described
\cite{meyerwallach} as a global entanglement measure. However, the
$MW$ measure precludes a detailed knowledge of the different
quantum correlations $I_{i_{1}i_{2}\ldots i_{k}}$ in the system.
For example, it cannot distinguish fully entangled states from
states that, in spite of being entangled, are separable in some of
their subsystems \cite{love}.

The generalized concurrence \cite{rungta} is a natural extension of
the concurrence (\ref{concqudit}) for multipartite N-qubit states
and can be expressed as \cite{andre}
\begin{equation}
\label{gconc}
C_{N}=2^{1-\left(  N/2\right)  }\sqrt{\left(
2^{N}-2\right) - {\sum_{\alpha}} \rm{tr}\rho_{\alpha}^{2}},
\end{equation}
where $\alpha$ labels all possible reduced density matrices. Note
that the expression inside the square-root is nothing more than
the sum of the concurrences in all possible bipartitions. The
generalized concurrence can detect real multipartite correlations
and it allows one to compare the degree of entanglement of
multipartite systems with different numbers of constituents, as
opposed to the $MW$ measure.

In the following section, we show that, for the states
(\ref{wstate}), (\ref{wlike}), (\ref{wlike0a}), and
(\ref{wlike0b}),  any convex (bipartite or multipartite)
entanglement quantifier that does not increase under LOCC's, in
any given partition, can be expressed in terms of that of a
considerably smaller subsystem consisting only of those qubits
lying on the boundary of the partition. No optimization on the
full system's parameter space is required throughout.

\section{Entanglement Dynamics of W states}\label{ed}

Following the same line of thought as in \cite{cavalcanti}, we
decompose the $W$ state as a set of two-qubit unitary
transformations acting on a separable state; the unitary
transformations that act inside the parts are local unitary
operations with respect to the partition and therefore do not
change its entanglement.

\begin{figure}
\begin{center}
\includegraphics*[width=0.7\linewidth]{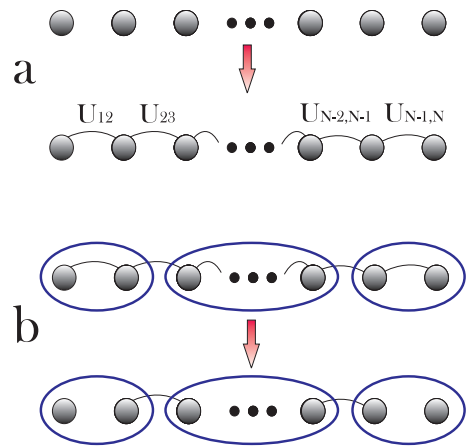}
\caption{(Color online) (a). Starting from a separable state the W
state is constructed by applying a series of unitary
transformations given by (\ref{uforw}). (b). The entanglement
corresponding to a given partition of the W state is equivalent to
the entanglement of a subsystem consisting of only the qubits
connected by the unitary transformations that cross the boundaries
of the partition.} \label{entanglementpartitions}
\end{center}
\end{figure}

To generate the states $\left\vert W_{N}\right\rangle $ and
$\left\vert W_{N}^{0}\right\rangle $ using two qubits operations
and starting with a separable state, we need to apply a series of
operations of the type 
\begin{equation}
U_{i,j}^{\mu,\nu}=
\begin{pmatrix}
1 & 0 & 0 & 0\\
0 & -\sqrt{\frac{\mu-\nu}{\mu}} & \sqrt{\frac{\nu}{\mu}} & 0\\
0 & \sqrt{\frac{\nu}{\mu}} & \sqrt{\frac{\mu-\nu}{\mu}} & 0\\
0 & 0 & 0 & 1
\end{pmatrix}
, \label{uforw}
\end{equation}
where $i,j$ are the qubits affected by the transformation, while
$\mu$ and $\nu$ are parameters defining the transformation. States
(\ref{wstate}) and (\ref{wlike0a}) can be written as [see Fig.
\ref{entanglementpartitions}(a)]
\begin{align}\label{wasu}
\left\vert W_{N}\right\rangle  &  =U\left\vert 10..00\right\rangle
,\nonumber\\
\left\vert W_{N}^{0}\right\rangle  &  =U\left(  \alpha\left\vert
10..00\right\rangle +\beta\left\vert 00..00\right\rangle \right),
\end{align}
with $U= {\prod_{i=0}^{N-2}} U_{N-1-i,N-i}^{i+2,i+1} $. Different
orderings of the operators $U_{i,j}^{\mu,\nu}$ are possible
depending on the kind of partition we are interested in. This is
explicitly shown in Eq. (\ref{wasdifferentu}). Also the
$\left\vert \widetilde{W}_{N}^{0}\right\rangle $ state can be
decomposed in this form, but now the two-qubit transformations
explicitly depend on the parameters $a_{i}$ that define the states
(\ref{wlike}) and (\ref{wlike0b}). This decomposition of the
states is now used to calculate their entanglement dynamics. 

\textit{Amplitude Damping:} The action of individual channels of
this type on the states (\ref{wstate}) yields 
\begin{equation}
\rho_{W}^{AD}=p\left\vert 0_{N}\right\rangle \left\langle 0_{N}\right\vert
+(1-p)\left\vert W_{N}\right\rangle \left\langle W_{N}\right\vert ,
\label{waddecohered}
\end{equation}
with $\left\vert 0_{N}\right\rangle =\left\vert 0\right\rangle
^{\otimes N}$. This state can be written in terms of the unitary
transformations (\ref{uforw}) as 

\begin{eqnarray}
\rho_{W}^{AD}  & =U[\left(  p\left\vert 00\right\rangle \left\langle
00\right\vert +\left(  1-p\right)  \left\vert 10\right\rangle
\left\langle
10\right\vert \right) \nonumber  \\
& \otimes\left\vert 0_{N-2}\right\rangle \left\langle
0_{N-2}\right\vert ]U^{\dagger}. \label{wdecoheredasu}
\end{eqnarray}

The states $\left\vert W_{N}\right\rangle $ and $\left\vert
W_{N}^{0}\right\rangle $ are invariant by permutations, so which
qubits are in each part is irrelevant. For the least balanced
bipartition $\{1\}:\{N-1\}$, only the unitary U$_{12}$ acting on
the boundary of this bipartition affects the entanglement, since
the unitary transformations acting inside the parts are local
transformations with respect this partition [see Fig.
\ref{entanglementpartitions}(b)]. So the entanglement in this case
can be seen to be 
\begin{align}\label{entanglement1xrestadreduced}
E(\rho_{W}^{AD})=E\left(  p\left\vert 0_{2}\right\rangle
\left\langle 0_{2}\right\vert +(1-p)\left\vert W_{2}(1)\right\rangle
\left\langle W_{2}(1)\right\vert \right)  ,
\end{align}
with $\left\vert W_{2}(k)\right\rangle =U_{1,2}^{N,N-k}\left\vert
10\right\rangle =\sqrt{\frac{N-k}{N}}\left\vert 01\right\rangle
+\sqrt {\frac{k}{N}}\left\vert 10\right\rangle $. The passage from
(\ref{wdecoheredasu}) to (\ref{entanglement1xrestadreduced}) follows
from the fact that the local addition of ancillas does not change
the entanglement. This is similar to the strategy adopted in
\cite{cavalcanti}, where it was shown that the entanglement dynamics
in a given partition of a graph state is the same as the one of the
particles in the boundary of the partition. For a bipartition
$\{k\}:\{N-k\}$ we can order the unitary transformations so that the
$W$ state is expressed as 
\begin{eqnarray}
\left\vert W_{N}\right\rangle &=&
{\displaystyle\prod\limits_{j=k}^{2}}
{\displaystyle\prod\limits_{i=k}^{N-2}}
\left(U_{j,j-1}^{j,j-1}U_{i+1,i+2}^{N-i,N-i-1}\right)\nonumber\\
&\times&U_{k,k+1}^{N,N-k}\left\vert
0_{1}\ldots1_{k}\ldots0_{N}\right\rangle . \label{wasdifferentu}
\end{eqnarray}

Following the same reasoning, the entanglement in this bipartition
$\{k\}:\{N-k\}$ can be seen to be
\begin{align}\label{entanglementkxNkadreduced}
E(\rho_{W}^{AD})=E\left(  p\left\vert 0_{2}\right\rangle
\left\langle 0_{2}\right\vert +(1-p)\left\vert W_{2}(k)\right\rangle
\left\langle W_{2}(k)\right\vert \right)  .
\end{align}

The same is valid for higher-order partitions. Depending on the
way the entangled state is partitioned, the order of the operators
$U_{i,j}^{\mu,\nu}$ must be properly chosen so as to factor out
the unitary transformations that are irrelevant for the
entanglement in the partition under consideration. For instance,
suppose we want to calculate the entanglement in a tripartition
$\{1\}:\{2\}:\{3,4\}$ of a four-qubit $W$ state. If one write this
state as $U_{12}U_{23}U_{34}$ applied to a separable state, then
every single unitary transformation must be taken into account, as
opposed to the alternative description $U'_{34}U'_{12}U'_{23}$
applied to another separable state, where the last unitary
transformation $U'_{34}$ can be neglected, since it is local with
respect to the part $\{3,4\}$.

This formalism is valid for any convex and monotonic entanglement
measure. Similar expressions can be found to (\ref{wlike}),
(\ref{wlike0a}), and (\ref{wlike0b}) evolving under the AD
channel.

\textit{Dephasing:} The action of individual channels of this type on the state (\ref{wstate}) leaves the diagonal elements untouched, while the coherences gain a factor (1-p)$^{2}$. The evolved state $\left\vert W_{N}\right\rangle $ can be written as%
\begin{equation}
\rho_{W}^{D}=(1-p^{\prime})
{\displaystyle\sum\limits_{k=1}^N}
\left\vert k_{N}\right\rangle \left\langle k_{N}\right\vert +p^{\prime
}\left\vert W_{N}\right\rangle \left\langle W_{N}\right\vert
,\label{dephasedW}
\end{equation}
with $p'=(1-p)^{2}$ and
$|k_N\rangle=(1/\sqrt{N})|0,\dots,1_k,\dots,0\rangle$. Following
the same recipe as for the $AD$ channel, the entanglement in a
bipartition $\{k\}:\{N-k\}$ of the dephased $\left\vert
W_{N}\right\rangle $ is given by 
\begin{eqnarray}
E(\rho_{W}^{D})&=&E\left(
(1-p^{\prime})
{\displaystyle\sum\limits_{k=1}^N}
\left\vert k_{N}\right\rangle \left\langle k_{N}\right\vert\right.\nonumber \\
&+&p^{\prime}\left\vert W_{2}(k)\right\rangle \left\langle W_{2}(k)\right\vert
\otimes\left\vert 0_{N-2}\right\rangle \left\langle 0_{N-2}\right\vert
\Bigg).\label{dephasedWusingU}
\end{eqnarray}
We apply to the last $N-2$ qubits in the state
(\ref{dephasedWusingU}) a separable POVM \cite{POVM separable} with
two elements given by $A_{1}={\sum_{k=1}^{N-2}}\left\vert
k_{N-2}\right\rangle \left\langle k_{N-2}\right\vert $ with $A_{1}^{\dag}%
A_{1}={\sum_{k=1}^{N-2}}\left\vert k_{N-2}\right\rangle
\left\langle k_{N-2}\right\vert <I_{N-2}$ and
$A_{2}=I_{N-2}-{\sum_{k=1}^{N-2}}\left\vert k_{N-2}\right\rangle
\left\langle k_{N-2}\right\vert $ with $A_{2}^{\dag }A_{2}=A_{2}$.
The completeness relation
$A_{1}^{\dag}A_{1}+A_{2}^{\dag}A_{2}=I_{N-2}$ is satisfied. The
post measurement state associated with $A_{1}$ is 
\begin{equation}
\frac{A_{1}\rho_{W}^{D}A_{1}^{\dag}}{\rm{tr}\left(
A_{1}\rho_{W}^{D}A_{1}^{\dag }\right) }={\sum_{k=1}^{N}}\left\vert
k_{N}\right\rangle \left\langle k_{N}\right\vert
\end{equation}
and the state associated with $A_{2}$ is
\begin{equation}
\frac{A_{2}\rho_{W}^{D}A_{2}^{\dag}}{\rm{tr}\left(
A_{2}\rho_{W}^{D}A_{2}^{\dag }\right)  }=\left\vert
W_{2}(k)\right\rangle \left\langle W_{2}(k)\right\vert
\otimes\left\vert 0_{N-2}\right\rangle \left\langle
0_{N-2}\right\vert\,.
\end{equation}
The POVM is separable and cannot raise the entanglement, which leads to
a lower bound for the entanglement
\begin{align}
E(\rho_{W}^{D})  & \geq p^{\prime}E\left(  \left\vert
W_{2}(k)\right\rangle \left\langle W_{2}(k)\right\vert \right)\,.
\nonumber
\end{align}

The convexity of entanglement, directly applied to
(\ref{dephasedW}), yields an upper bound for the entanglement
\begin{equation}
E(\rho_{W}^{D})\leq p^{\prime}E\left(  \left\vert
W_{2}(k)\right\rangle
\left\langle W_{2}(k)\right\vert \right)  .\label{dephasedupperbound}%
\end{equation}
The lower and upper bounds coincide and therefore give an exact
quantification of the entanglement. Similar expressions can be
found for the states (\ref{wlike}), (\ref{wlike0a}), and
(\ref{wlike0b}) undergoing a dephasing process.

\section{Bipartite versus Global Entanglement Dynamics}\label{bg}

Using as a measure of nonlocal information the $n$-tangle
\cite{Ntangle}, we show in the Appendix that $W$-like states
(\ref{wstate}), (\ref{wlike}), (\ref{wlike0a}), and
(\ref{wlike0b}) are completely characterized by $2$-tangles
\begin{equation}
E_{MW}\left(\left\vert
\widetilde{W}_{N}^{0}\right\rangle\right)=\frac{2}{N}\sum_{i_{1}<i_{2}}\tau\left(
\left\vert \widetilde{W}_{N}^{0}\right\rangle \right)
_{i_{1}i_{2},}\,, \label{wonly2tangle}
\end{equation}
all the other genuine multipartite entanglement tangles being
null. Since the decoherence acting on the states is local, which
cannot increase the entanglement,  the global entanglement
dynamics is completely specified by the $2$-tangle dynamics. This
is a fortunate circumstance, since the $2$-tangle can be
analytically evaluated even for mixed states, being directly
related to the concurrence, while higher-order tangles require, in
general, involved optimizations.

For the two-qubit state $\rho_{ij}$ obtained after tracing over
$N-2$ qubits, the $2$-tangle for the AD channel is given by
$\tau_{ij}^{AD}=\left\vert \alpha\right\vert ^{4}\left\vert
a_{i}\right\vert ^{2}\left\vert a_{j}\right\vert ^{2}(1-p)^{2}$,
while for the dephasing channel we get $\tau_{ij}^{D}=\left\vert
\alpha\right\vert ^{4}\left\vert a_{i}\right\vert ^{2}\left\vert
a_{j}\right\vert ^{2}(1-p)^{4}$. The $MW$ measure, in this case
the sum over all $2$-tangles, is therefore given, respectively, by
\begin{align}
E_{MW}(\rho^{AD})  & =\frac{2}{N}(1-p)^{2}\sum_{i<j}^{N}\left\vert
\alpha\right\vert ^{4}\left\vert a_{i}\right\vert ^{2}\left\vert
a_{j}\right\vert ^{2}\label{MWAD}\\
& =(1-p)^{2}E_{MW}\left(\left\vert
\widetilde{W}_{N}^{0}\right\rangle \right),\nonumber
\end{align}
and
\begin{equation}
E_{MW}(\rho^{PD})=(1-p)^{4}E_{MW}\left(\left\vert \widetilde{W}_{N}^{0}%
\right\rangle\right). \label{MWPD}%
\end{equation}
In both cases, the $MW$ entanglement decay is size independent. This
is a generalization of the result obtained in \cite{montakhab},
restricted to the state (\ref{wstate}).

One should note that our method, as proposed in Sec. \ref{ed},
allows a considerable simplification in the evaluation of the
entanglement corresponding to a given bipartition. Indeed, in the
usual approach, the negativity associated to some bipartition is
calculated by partially transposing the density matrix. Given the
symmetry of the states (\ref{wstate}) and (\ref{wlike0a}) under
permutation of the qubits, we can restrict these partial
transpositions to the $N/2$ first qubits ($N/2+1$ for $N$ odd).
Under the partial transposition on the bipartition $\{k\}:\{N-k\}$
of the evolved density matrix, $2k(N-k)$ coherences, originally in
the subspace of one excitation, go to the subspace of two
excitations. With this remark we could in principle calculate
analytically the eigenvalues of the partially transposed density
matrix as functions of $p$, $N$, and $k$, which would involve in
general finding the roots of a  $N^{2}$th order polynomial.
However, as we have shown in the last section, the bipartite
entanglement can be calculated through a much simpler way, using a
two-qubit density matrix.

Using the concurrence as the quantifier of bipartite entanglement
we see even more clearly the power of our method. For multipartite
mixed states, the calculation of the concurrence would be
attempted by a brute-force optimization approach given by
Eq.~(\ref{concmixed}), but since, in this particular case we treat
here, the boundary qubits are just two, the use of our method
allows us to perform the calculation with no optimization at all,
for an explicit formula for the concurrence exists for arbitrary
two-qubit systems.

For the state (\ref{wstate}) undergoing dephasing and amplitude
damping, the negativities in a bipartition $\{k\}:\{N-k\}$ can be
shown to be given, respectively, by 
\begin{equation}
{\cal N}_{D}(p,N,k)=\frac{\left(  1-p\right)  ^{2}}{N}\sqrt{k\left(  N-k\right)
}\text{,} \label{negdephasing}
\end{equation}
\begin{equation}
{\cal N}_{AD}(p,N,k)=-\frac{p}{2}+\frac{1}{2N}\sqrt{N^{2}p^{2}+4k(N-k)(1-p)^{2}}.
\label{negad}
\end{equation}

The concurrences for bipartitions $\{k\}:\{N-k\}$ of the state
(\ref{wstate}) undergoing dephasing and amplitude damping are
given, respectively, by 
\begin{equation}
C_{D}(p,N,k)=2\frac{\left(  1-p\right)  ^{2}}{N}\sqrt{k\left(
N-k\right) }\text{,} \label{concdephasing}
\end{equation}
\begin{equation}
C_{AD}(p,N,k)=2\frac{\left(  1-p\right) }{N}\sqrt{k\left( N-k\right)
}\text{,} \label{concad}
\end{equation}

\begin{figure}[!t]
\begin{center}
\includegraphics*[width=0.7\linewidth]{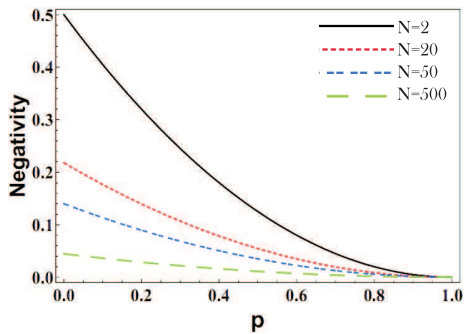}
\caption{(Color online) The negativity corresponding to the least
balanced bipartition as a function of $p$ and $N=2,20,50,500$, for
the W state undergoing individual dephasing. The initial
entanglement decreases as the number of qubits increases, but the
decay factor is independent of $N$ and given by $(1-p)^2$.}
\label{initialneg}
\end{center}
\end{figure}

Setting $p=0$ we can analyze the bipartite entanglement of the
initial $W$ state. For $k=N/2$ the entanglement is independent of
$N$ and maximal; in fact the balanced bipartition can always be
written in terms of effective qubits as a maximally entangled
two-qubit state $\left\vert \Psi^{+}\right\rangle
=\frac{1}{\sqrt{2}}\left( \left\vert 01\right\rangle +\left\vert
10\right\rangle \right)$. For other $k$'s the initial bipartite
entanglement of the $W$ state depends on $N$; for a given value of
$k$ the $W$ state behaves like an effective two-qubit entangled
state of the form $\left\vert \widetilde{\Psi^{+}}\right\rangle
=\left( \sqrt{\frac{N-k}{N}}\left\vert 01\right\rangle
+\sqrt{\frac{k}{N}}\left\vert 10\right\rangle \right)  $, for which
the initial entanglement has an intrinsic dependence on $N$. For a
fixed $k$, the larger $N$ is the more this bipartition approximates
a separable one. This can be viewed in Fig. \ref{initialneg}, where
we plot the negativity of the least balanced bipartition as a
function of time in the state (\ref{wstate}) undergoing dephasing.
The initial entanglement decreases with the number of qubits.

From Eqs.~(\ref{negdephasing}), (\ref{negad}),
(\ref{concdephasing}), and (\ref{concad}) we can see that the $W$
state does not undergo finite-time disentanglement. However, if we
are interested not in the disappearance of the initial
entanglement, but in the survival of a significant fraction of it,
either to be directly used or to be distilled without an
excessively large overhead in resources, we need to look at its
decay \cite{aolita}, that is, the behavior of the ratios
$\frac{{\cal N}(p,N,k)}{{\cal N}(0,N,K)}=\epsilon$ and
$\frac{C(p,N,k)}{C(0,N,K)}=\delta$ as a function of $p$.

Under dephasing the decay of negativity and concurrence of any
bipartition is expressed by the factor
\begin{equation}
\epsilon_{D}=\delta_{D}=\left(  1-p\right)  ^{2}.
\label{decaydephasing}
\end{equation}

\begin{figure}[!t]
\begin{center}
\includegraphics*[width=0.7\linewidth]{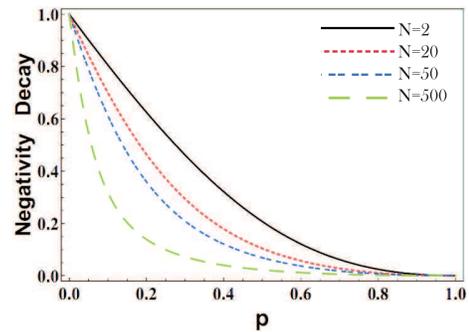}
\caption{(Color online) The decay of the negativity corresponding to
the least balanced bipartition as a function of $p$ for
$N=2,20,50,500$, for the $W$ state undergoing individual amplitude
damping. Even though the global entanglement decay in the state is
size independent, the entanglement decay in this bipartition, for a
fixed value of $p$, scales as  $1/\sqrt{N}$, for $N$ sufficiently
large. } \label{addecay}
\end{center}
\end{figure}

For the amplitude damping the concurrence decay for any bipartition
is size independent and given by 
\begin{equation}
\label{deltaad} \delta_{AD}=\left(  1-p\right) .
\end{equation}
while the negativity decay is given by
\begin{equation}
\epsilon_{AD}(p,N,k)=\frac{-Np+\sqrt{N^{2}p^{2}+4k(N-k)(1-p)^{2}}}
{2\sqrt{k(N-k)}}, \label{ad-decay}
\end{equation}
which is independent of $N$ for $k=N/2$. However, the more
unbalanced is the bipartition, the faster is its decay. The idea
is clearly illustrated in Fig. (\ref{addecay}) for the least
balanced bipartition. For $k=1$ and any $p\ne0$, it is easy to see
that
\begin{equation}
\epsilon_{AD}(p,N,1)\propto {1\over \sqrt{N}}\quad {\rm when}\quad
N>>\left\{1, {(1-p)^2\over p^2}\right\}\,,
\end{equation}
which contrasts with the behavior of the $GHZ$ states under the same
channel, where for a fixed value of $p$ the negativity decays
exponentially with $N$ \cite{aolita}.

The size independence for the decay factor for the concurrence in
any bipartition of a W state (\ref{wstate}) is also reflected in a
size independence for the decay of the generalized concurrence.
This can be easily seen through Eq.~(\ref{gconc}), since the decay
factors, expressed in Eqs.~(\ref{decaydephasing}) and
(\ref{deltaad}), are independent of the bipartition. This is the
analytical proof of the result obtained by numerical methods
in~\cite{andre}.

Under dephasing the bipartite entanglement decay is independent of
$N$ and $k$, the same being true for the global entanglement
\cite{andre,montakhab}. The $W$ state is extremely robust under
the dephasing channel. Under the action of the amplitude damping
channel, the bipartite entanglement of $W$ states is much more
robust than that of $GHZ$ states, although it can still depends
intrinsically on the number of particles $N$. This is an
interesting and unexpected behavior since the global entanglement
decay, as quantified by the $MW$ measure and the generalized
concurrence, is independent of $N$. The same conclusions are valid
for the states (\ref{wlike}), (\ref{wlike0a}), and
(\ref{wlike0b}): under dephasing and amplitude damping the global
entanglement decay is size independent while the negativity decay
can depend on $N$ for the $AD$ channel.

The different behavior of the global and bipartite
quantifiers is now shown to have direct implications on the
robustness of two distinct communication tasks. 

\section{Teleporting and Splitting Quantum Information}\label{teleport}

Entangled states are resources for important protocols for quantum
communication. Let us consider now the effect of decoherence on two
related tasks, which we show to be associated with different kinds
of entanglement.

We consider first the teleportation \cite{teleportation} of an
unknown quantum state $\left\vert \psi\right\rangle = {\sum_{i}^{d}}
a_{i}\left\vert i\right\rangle $ of dimension $d$. To perfectly
realize the teleportation, Alice and Bob need to share a maximally
entangled bipartite state $\left\vert \Phi^{+}\right\rangle
_{AB}=\left(1/\sqrt{d}\right) {\sum_{i}^{d}} \left\vert
i\right\rangle _{A}\left\vert i\right\rangle _{B}$. If they share a
non-maximally entangled state they will be able to realize an
imperfect teleportation. The quality of the teleportation is
quantified by the fidelity, defined as
$f=\langle\psi|\rho|\psi\rangle$, where $\rho$ is the state obtained
after teleportation. The maximum achievable fidelity $f_{\max}$ for
a given bipartition of the state used as a channel is bounded by the
negativity $\mathcal{N}$ associated to this bipartition
\cite{teleportation, vidal}:
\begin{equation}
f_{\max}\leq\frac{2+2\mathcal{N}}{d+1}. \label{teleportationbound}
\end{equation}

The splitting of quantum information \cite{splitting} is a
generalization of the teleportation protocol. Alice has an unknown
qubit $\left\vert \Psi_{0}\right\rangle =\cos\left(
\theta/2\right) \left\vert 0\right\rangle +e^{i\phi}\sin\left(
\theta/2\right)  $ $\left\vert 1\right\rangle $ she would like to
send to $N$ other parts, a many-Bobs system,  in such a way that
the $N$ Bobs must cooperate in order that just one of them
extracts the quantum information. This is the best one can have,
in view of  the no-cloning theorem \cite{nocloning}, which implies
that only one copy of $\left\vert \Psi_{0}\right\rangle $ can be
received. Alice and the other $N$ parts share an entangled state
and proceed in a very similar way as in the usual quantum
teleportation \cite{teleportation}. First, Alice teleports the
qubit to the $N$ Bobs, a usual teleportation related to the
negativity in the bipartition. We show now that the second part of
the protocol, when the $N$ parts cooperate to extract the
information, is related to the global entanglement of the shared
state.

We consider, under decoherence, two distinct protocols to split
quantum information, which use $GHZ$ and a $W$-like state as
resources. Note from (\ref{teleportationbound}) that to teleport a
qubit we need to use a channel that has at least
$\mathcal{N}=1/2$. This is true for any bipartition of a $GHZ$
state \cite{aolita}, but as we have shown this is no longer true
for any bipartion of a $W$ state (\ref{wstate}). The only
bipartition of a $W$ state with $\mathcal{N}=1/2$ is the balanced
bipartition. To use a $W$ state as the channel for the splitting
of information between the $N$ Bobs, we need to use the balanced
partition in a $W$ state with $2N$ qubits. The more unbalanced is
the bipartition the worse is the initial teleportation fidelity
and the faster will be the fidelity decay under decoherence.
Another possible choice for a perfect teleportation, the one we
consider here, is to use an asymmetric $W_{A}$ state with $N+1$
qubits of the form 
\begin{equation}
\left\vert W_{A}^{N+1}\right\rangle =\frac{1}{\sqrt{2}}\left[
\left\vert 0\right\rangle\left\vert W_{N}\right\rangle  +\left\vert
1\right\rangle\left\vert 0\right\rangle^{\otimes N} \right]  ,
\label{assymetricw}
\end{equation}
which has the required $\mathcal{N}=1/2$ in the bipartition
$\{1\}:\{N\}$.

\begin{figure}
\begin{center}
\includegraphics*[width=0.7\linewidth]{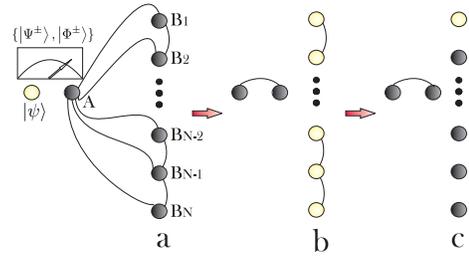}
\caption{(Color online) Scheme for the teletransport and the
splitting of quantum information involving $N+1$ parts. The edges
represent the entanglement between the qubits, which is created or
consumed along the protocol. (a). Alice measure her qubit and the
qubit $\left\vert \psi\right\rangle$ she wants to teletransport
(split) in a Bell basis. (b). The teletransported state is encoded
in a $N$ qubit state. (c). The $N$ parties must cooperate so that
only one of them has at the end the teleported state $\left\vert
\psi\right\rangle$. } \label{scheme}
\end{center}
\end{figure}

In both the $GHZ$ and $W_{A}$ protocols, Alice measures the two
qubits in her possession in a Bell basis, communicating the
classical outcomes to the Bobs (Fig. \ref{scheme}). In the $GHZ$
protocol, after the Bell measurement performed by Alice, the $N$
Bobs state will be given by one of the four states 
\begin{align}
\left\vert \widetilde{\Phi}^{+.-}\right\rangle  &  =\alpha\left\vert
0\right\rangle ^{\otimes N}\pm\beta\left\vert 1\right\rangle
^{\otimes
N},\label{ghzdecodification}\\
\left\vert \widetilde{\Psi}^{+.-}\right\rangle  &  =\alpha\left\vert
1\right\rangle ^{\otimes N}\pm\beta\left\vert 0\right\rangle
^{\otimes N}.\nonumber
\end{align}
After they decide whom among them will be the receiver of the
state $\left\vert \Psi_{0} \right\rangle $,  the
non-receivers measure their qubits in the basis of  eigenvectors
of $X$, where $X$ is the $\sigma_x$ Pauli matrix, and communicate their outcomes to the receiver. In possession
of all the classical measurement outcomes, the receiver can recover
the teleported state, since the state in his possession is
\begin{equation}
Z^{M}X^{a_{2}}Z^{a_{1}}\left\vert \Psi_{0}\right\rangle ,
\end{equation}
$M$ being the number of Bobs measurements that returned
$\left\vert -\right\rangle $, and the indices a$_{1}$ and a$_{2}$
are related to the measurement made by Alice, so that: $\left\{
\left\vert \Phi^{+}\right\rangle ,\left\vert \Psi^{+}\right\rangle
\right\} \Rightarrow a_{1}=0$, $\left\{ \left\vert
\Phi^{-}\right\rangle ,\left\vert \Psi^{-}\right\rangle \right\}
\Rightarrow a_{1}=1$, $\left\{ \left\vert \Phi^{+}\right\rangle
,\left\vert \Phi^{-}\right\rangle \right\} \Rightarrow a_{2}=0$,
and $\left\{ \left\vert \Psi^{+}\right\rangle ,\left\vert \Psi
^{-}\right\rangle \right\} \Rightarrow a_{2}=1$.

Using the $W_{A}$ state as the resource, after Alice's measurement
the four possible $N$-qubit state outcomes are 
\begin{align}
\left\vert \widetilde{\Phi}^{+.-}\right\rangle  &  =\alpha\left\vert 0\right\rangle ^{\otimes N} \pm \beta  \left\vert
W_{N}\right\rangle , \label{assymetricaftermeasurement}\\ \left\vert \widetilde{\Psi}^{+.-}\right\rangle  &  = \alpha  \left\vert W_{N}\right\rangle   \pm\beta\left\vert 0\right\rangle ^{\otimes N} .\nonumber
\end{align}
Note this is nothing more than the usual teleportation, with the
information encoded in effective qubits. Note also that the shared
state is in the form (\ref{wlike0a}). In the decodification part,
all the $N$ Bobs need to meet, and then apply a global operation on
their qubits, so that 
\begin{align}
\left\vert W_{N}\right\rangle   & \rightarrow\left\vert 1...00\right\rangle ,
\label{decodification}\\
\left\vert 0...00\right\rangle  &  \rightarrow\left\vert 0...00\right\rangle ,\nonumber
\end{align}
in such a way the information is now only contained in one of the
qubits \cite{splitting}. The above transformation is realized by
the inverse of the unitary transformation displayed in
Eq.~(\ref{wasu}).

Under amplitude damping of each qubit in the resource states $GHZ$
and $W_A$, the respective fidelities in the teleportation part of
the protocol, averaged over all the possible input states
$\left\vert \Psi_{0} \right\rangle $, can be shown to be

\begin{eqnarray}
\overline{f^{T}}_{GHZ}  &=&\frac{1}{6} \left[ 2+\left(  1-p\right)^{N-1}(2-p) \right. \\
&+&\left. 2\left(  1-p\right)  ^{N/2}+p^{N-1}(1+p) \right] \nonumber \\
\overline{f^{T}}_{W_{A}} &=&\frac{1}{3}\left(  3-2p+p^{2} \right),
\label{fidelities}
\end{eqnarray}

\begin{figure}[t]
\begin{center}
\includegraphics*[width=0.7\linewidth]{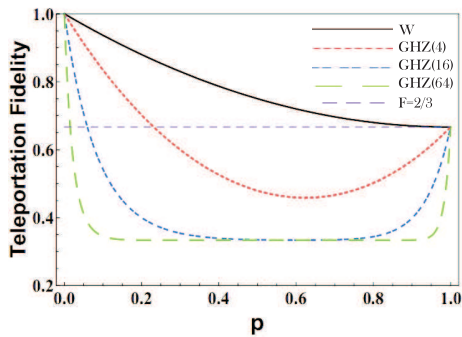}
\caption{(Color online) Teleportation fidelities for protocols that
use as resources $GHZ$ and $W_{A}$ states, under the $AD$ channel.
The $W_{A}$ protocol is the best possible under $AD$ and throughout
its evolution always leads to fidelities above the classical
threshold, while the $GHZ$ protocol  leads to fidelities below the
classical threshold.} \label{fidelitiesfig}
\end{center}
\end{figure}

 Considering the
complete splitting protocol, the teleportation followed by the
decodification, the fidelities are 
\begin{align}
\overline{f}_{GHZ}^{D} &  = \frac{1}{3}\left[  2-p\left(  1-p\right)  +\left(  1-p\right)  ^{N/2}\right] \nonumber\\
\overline{f^{D}}_{W_{A}} &  =1-\frac{p}{3}.
\end{align}
For the asymptotic separable state $(p=1)$ we see that the average
classical fidelity $2/3$ is recovered in all the previous
expressions. In the $W_{A}$ state, the entanglement corresponding
to the partition between Alice and the $N$ Bobs decays as given in
Eq.~(\ref{ad-decay}) with $k=N/2$.  The negativity in the
bipartition used for the teleportation in the state
(\ref{assymetricw}) under $AD$ is such that $f_{\max}$, as given
by Eq.~(\ref{teleportationbound}), is equal to
$\overline{f^{T}}_{W_{A}}$. This implies that not only is  the
protocol robust, but also it is the best possible. The same is not
true for the $GHZ$ protocol under decoherence, since
$\overline{f^{T}}_{GHZ}$ decays exponentially with $N$ and along
the evolution this fidelity can be under the classical limit, in
spite of the fact that the negativity is null only in the
asymptotic limit $p=1$. The teleportation fidelities are plotted
in Fig.~(\ref{fidelitiesfig}). 

The resource state used for the decodification is given by
(\ref{assymetricaftermeasurement}), which is a $W$-like state
(\ref{wlike0a}), which has a global entanglement decay independent
of $N$, while its bipartite entanglement, quantified by the
negativity, depends on $N$. The fact that the fidelity of the
whole protocol is size independent is a clear indication that the
decodification part of the protocol depends on the global
entanglement, which should be expected since this decodification
depends on all parts in a symmetric way. Therefore, the comparison
of the decays of the entanglement and of the fidelity indicates on
which kind of entanglement the task relies. In the decoherent
scenario, similar but non-identical protocols can rely on
different types of entanglement, which can imply distinct
robustness of these distinct tasks. For the teleportation and
splitting of quantum information, protocols based on $W$-like
states are clearly much more robust than $GHZ$-based protocols.

\section{Conclusions}\label{conc}

The characterization of entanglement for multipartite systems is a
complex endeavor, in view of the many possible partitionings of the
state and the possibility of having, for a given partition, many
possible quantifiers. Even for two qubits, two widely used
quantifiers, the concurrence and the negativity, do not lead to
consistent results when comparing the entanglement of two states.
This motivates the search for criteria that would associate each
quantifier with a different physical task.

This problem becomes even more pressing when considering the effects
of the environment. Different quantifiers may display quite distinct
behaviors under decoherence. This motivates a natural question: can
these different dynamics be associated to the robustness of
inequivalent communication tasks?

In this paper, we have shown that two of those quantifiers,
proposed for estimating global and bipartite entanglements,
respectively, do behave quite distinctively under decoherence, and
that this divergent behavior is directly related to the robustness
of two inequivalent communication tasks.

We have made a detailed comparison between the use of $GHZ$ or $W$
states as resources for quantum communication in a noisy
environment, by applying a new technique that allows a reduction of
the $W$-state entanglement dynamics to that of a two-qubit system.
This technique could be possibly applied to other classes of
entangled states under different kinds of environment.

Our approach suggests that the eventual ambiguities in the
definition of entanglement quantifiers should be resolved by
associating each quantifiers with a definite physical task.

\appendix
\section{W states have only $2$-tangle}
In order to show that (\ref{wonly2tangle}) is valid for the $W$-like
states $\left\vert \widetilde{W}_{N}^{0}\right\rangle$, we employ
the $MW$ measure in the form proposed in \cite{meyerwallach},
\begin{equation}
E_{MW}(\left\vert \psi\right\rangle )=\frac{4}{N}
{\displaystyle\sum\limits_{i=1}^{N}} D(l_{i}(0)\left\vert
\psi\right\rangle ,l_{i}(1)\left\vert \psi\right\rangle
).\label{globaloperational}
\end{equation}
Here l$_{i}(b)$ is a $\left(  C^{2}\right)  ^{\otimes
N}\rightarrow\left( C^{2}\right)  ^{\otimes N-1}$ map 
\begin{equation}
l_{i}(b)\left\vert k_{1},\ldots,k_{N}\right\rangle
=\delta_{bk_{i}}\left\vert
k_{1},\ldots,\widehat{k}_{i},\ldots,k_{N}\right\rangle ,
\label{functionl}
\end{equation}
where $\widehat{k}_{i}$ means absence of the term $k_{i}$ and
$\left\vert k_{1},...,k_{N}\right\rangle $ is the computational
basis with $k_{i}$=$\left\{ 0,1\right\}$. So when l$_{i}(b)$ acts
on an N-dimensional vector state $\left\vert \psi\right\rangle$,
it generates an (N-1)-dimensional vector state $\left\vert
\psi^{i,b}\right\rangle = {\sum_{k=1}^{2^{N-1}}}
u_{k}^{i,b}\left\vert k\right\rangle $, being k=$\left(
k_{1},...,k_{N-1} \right)  $. The quantity $D(l_{i}(0)\left\vert
\psi\right\rangle ,l_{i}(1)\left\vert \psi\right\rangle )$ is a
distance defined by 
\begin{equation}
D(l_{i}(0)\left\vert \psi\right\rangle ,l_{i}(1)\left\vert
\psi\right\rangle )= {\displaystyle\sum\limits_{x<y}} \left\vert
u_{x}^{i,0}u_{y}^{i,1}-u_{y}^{i,0}u_{x}^{i,1}\right\vert ^{2}.
\label{distance}
\end{equation}

The application of the operator $l_{i}$ to the $\left\vert
\widetilde{W}_{N}^{0}\right\rangle $ gives 
\begin{align}
l_{i}(0)\left\vert \widetilde{W}_{N}^{0}\right\rangle  &  =
{\displaystyle\sum\limits_{k=1}^{2^{N-1}}}
u_{k}^{i,0}\left\vert k\right\rangle =\beta\left\vert 0\right\rangle
+\sum_{j\neq i}^{N}\alpha a_{j}\left\vert j\right\rangle ,\\
l_{i}(1)\left\vert \widetilde{W}_{N}^{0}\right\rangle  &  =
{\displaystyle\sum\limits_{k=1}^{2^{N-1}}}
u_{k}^{i,1}\left\vert k\right\rangle =\alpha a_{i}\left\vert 0\right\rangle
.\nonumber
\end{align}
Therefore
\begin{eqnarray}
&D(l_{i}(0)\left\vert \widetilde{W}_{N}^{0}\right\rangle ,l_{i}(1)\left\vert
\widetilde{W}_{N}^{0}\right\rangle )  =
{\displaystyle\sum\limits_{x<y}}
\left\vert u_{x}^{i,0}u_{y}^{i,1}-u_{y}^{i,0}u_{x}^{i,1}\right\vert ^{2}\nonumber\\
&  =\sum_{j\neq i}^{N}\left\vert u_{j}^{i,0}u_{0}^{i,1}\right\vert ^{2}=
{\displaystyle\sum\limits_{j\neq i}^{N}}
\left\vert \alpha\right\vert ^{4}\left\vert a_{j}a_{i}\right\vert
^{2},
\end{eqnarray}
and the $MW$ entanglement measure for $\left\vert
\widetilde{W}_{N}^{0}\right\rangle$ is thus 
\begin{eqnarray}\label{mwa}
E_{MW}\left(  \left\vert \widetilde{W}_{N}^{0}\right\rangle \right)  &=\frac
{4}{N}
{\displaystyle\sum\limits_{i=1}^{N}}
{\displaystyle\sum\limits_{j\neq i}^{N}}
\left\vert \alpha\right\vert ^{4}\left\vert a_{j}a_{i}\right\vert ^{2}\nonumber\\
&=\frac{8\left\vert \alpha\right\vert ^{4}}{N}
{\displaystyle\sum\limits_{i<j}^{N}}
\left\vert a_{j}a_{i}\right\vert ^{2}.
\end{eqnarray}
Tracing out, in the state $\left\vert
\widetilde{W}_{N}^{0}\right\rangle$, every qubit but qubits $i$ and
$j$, the two-qubit density matrix $\rho_{ij}$ becomes 
\begin{equation}
\begin{pmatrix}
\left\vert \beta\right\vert ^{2}+\left\vert \alpha\right\vert ^{2}\left(
1-\left\vert a_{i}\right\vert ^{2}-\left\vert a_{j}\right\vert ^{2}\right)  &
\alpha\beta^{\ast}a_{i}^{\ast} & \alpha\beta^{\ast}a_{j}^{\ast} & 0\\
\alpha^{\ast}\beta a_{i} & \left\vert \alpha\right\vert ^{2}\left\vert
a_{i}\right\vert ^{2} & \left\vert \alpha\right\vert ^{2}a_{i}^{\ast}a_{j} &
0\\
\alpha^{\ast}\beta a_{j} & \left\vert \alpha\right\vert ^{2}a_{i}a_{j}^{\ast}
& \left\vert \alpha\right\vert ^{2}\left\vert a_{j}\right\vert ^{2} & 0\\
0 & 0 & 0 & 0
\end{pmatrix}
.
\end{equation}
The concurrence of this state is given by $C_{ij}=2\left\vert \alpha
\right\vert ^{2}\left\vert a_{i}a_{j}\right\vert $ and so the 2-tangle is
$\tau_{ij}=C_{ij}^{2}=4\left\vert \alpha\right\vert ^{4}\left\vert a_{i}%
a_{j}\right\vert ^{2}$.  From this result and from (\ref{mwa}) it can be easily seen that
\begin{equation}
E_{MW}(\left\vert \widetilde{W}_{N}^{0}\right\rangle
)=\frac{2}{N}\sum _{i<j}^{N}\tau_{ij}.
\end{equation}
Therefore, the global entanglement of state $\left\vert
\widetilde{W}_{N}^{0}\right\rangle $ is  completely characterized by
its $2$-tangles.

\textbf{Acknowledgements}.-- This work was supported by the
Brazilian agencies FAPERJ, CAPES, CNPq,  and the Brazilian National
Institute for Quantum Information.

\end{document}